\documentclass[11pt]{article}

\usepackage{amsfonts}
\usepackage{amsbsy} 
\input amssym.def
\input amssym.tex
\begin{document}

\newcommand{\half}{{1\over 2}}
\newcommand{\ben}{\begin{equation}}
\newcommand{\een}{\end{equation}}

\newcommand{\hk}{hyperK\"{a}hler}
\newcommand{\ka}{K\"{a}hler}
\newcommand{\ein}{Einstein}
\newcommand{\ma}{Monge--Amp\`ere}
\newcommand{\ti}[1]{\ensuremath{\tilde{#1}}}
\newcommand{\bs}{\ensuremath{\boldsymbol{\omega}}}
\newcommand{\re}{\ensuremath{\mathbb{R}}}
\newcommand{\co}{\ensuremath{\mathbb{C}}}
\newcommand{\q}{\ensuremath{\mathbb{H}}}
\newcommand{\z}{\ensuremath{\mathbb{Z}}}
\newcommand{\euc}{\ensuremath{\mathbb{E}}}
\newcommand{\p}{\ensuremath{\mathbb{P}}}
\hfuzz=100pt
\title{Single-sided domain walls in M-theory} 
 \author{G.~W.~Gibbons\thanks{e-mail:gwg1@damtp.cam.ac.uk}\ \  and \    
        P.~Rychenkova\thanks{e-mail:pr201@damtp.cam.ac.uk} \\
        DAMTP, University of Cambridge, Silver Street \\ 
        Cambridge, CB3 9EW, U.K.} 
\date{November 4, 1998}
\maketitle
\medskip
\centerline{DAMTP-1998-149}

\begin{abstract} We describe some single-sided BPS domain wall
configurations in M-theory. These are smooth non-singular resolutions
of Calabi--Yau orbifolds obtained by identifying the two  sides of the
wall under reflection. They may thus be thought of as domain walls at
the end of the universe. We also describe related domain wall type
solutions with a negative cosmological constant.
\end{abstract}

\section{Introduction} \label{intro}

It is now widely recognized that topological
defects with $p$ spatial dimensions   invariant under  half the maximum
number of supersymmetries --- BPS configurations ---
 play a central role in non-perturbative string theory and M-theory. If $p$
is less than $(n-3)$, where $n$ is the space-time dimension, then such objects
can be studied in at least two limits.  One is the {\it light} approximation
in which 
the gravitational field that the objects generate is ignored. Treated
classically,  the world-volume theory of such objects is described by  
a Dirac--Born--Infeld type action. The other is the {\it heavy}
approximation, in which the gravitational field generated by the
objects is taken into
account and one looks for solutions of the supergravity equations of motion. 

If $p< n-3\ $ heavy branes  give rise to  asymptotically flat  metrics
in directions transverse to the brane,
 and from a distance they behave more
or less like light branes moving in a flat background.  If $p=n-3\ $
(vortices)  or $p=n-2\ $ (domain walls),
 however, the metrics they generate are not asymptotically flat.
 For vortices --- for example, the 7-brane of 
 the ten-dimensional type IIB
 theory, --- the metric has an angular deficit.
 In the case of domain walls, their effect on space-time can be even
 more drastic. 
 For example,  conventional domain walls, even in the thin-wall
approximation,  bring about the compactification of space
\cite{gibbons93}.  This happens as
follows. Either side of the domain wall is isometric to the interior of a
time-like hyperboloid in Minkowski spacetime $\euc^{\, n-1,1}$. To get the
entire spacetime one glues two such domain walls back to back. The induced
metric is continuous across the domain wall but the second fundamental form
has a discontinuity which gives the distributional stress tensor.
Another feature of conventional domain walls, which is more or less
obvious from the description just given, is that one does
not expect to have more than one in a static configuration.

Domain walls of an unconventional
(orbifold) type play an important role in Ho\v{r}ava and Witten's 
approach to the $E_8 \times E_8$ heterotic
string theory in M-theory \cite{horavawitten1,horavawitten2}. They also have a
drastic global effect
on the structure of space-time. Of course in addition to their gravitational
fields one must take into account the effects of anomalies
and the four-form field strength.

In this paper we are going to study the global structure of  
some other spacetimes containing BPS domain walls that have arisen in
M-theory. A striking feature of M-theory is the extent to which
configurations in eleven dimensions are non-singular even though they may
appear to be singular in lower dimensions. We shall therefore be 
particularly interested in everywhere non-singular configurations.
The organization of the paper is summarized by the contents table.
\tableofcontents

\section{Bianchi domain walls} \label{bianchisect}

We shall consider  $p$-brane solutions of the form:
\[
M_4 \times \euc^{\, p-3,1},
\]
where $M_4$ is a Riemannian four-manifold which is either Ricci-flat
or has negative cosmological constant.
If $p=3$ we would be considering domain walls in five space-time
dimensions. We are looking for metrics on $M_4$ which depend only on
one coordinate $t$, transverse to the domain wall. The metric should
be homogeneous in the directions parallel to the wall. Mathematically
this means that we are looking for cohomogeneity one, or hypersurface
homogeneous, metrics invariant under the action of  Lie group $G$
acting transitively on three-dimensional orbits. In the cases we are
interested in $G$ may be taken to be three-dimensional and the
possible groups have been classified by Bianchi (see
e.g. \cite{lifshitz}). The problem is very similar to that encountered
in studying homogeneous Lorentzian cosmologies and we shall freely use
standard results from that subject \cite{siklos}.
 The Bianchi types relevant to this paper are type I, II,
VI$_0$ and VII$_0$ .  Domain walls of types
I and II are discussed in section~\ref{vacuumsect} while the treatment
of the  more
``exotic''  solutions is relegated to section~\ref{exoticsect}. 

In the following we shall find that all Ricci-flat solutions are
singular  and describe
how the singularity of the type II solution may be resolved. The
resolution of this singularity gives a complete Ricci-flat \ka\ manifold
that we shall call the BKTY metric.\footnote{The name BKTY is derived
  from the initials of the authors of
\cite{tianyau89,tianyau90,bandokobayashi88,bandokobayashi90} who
constructed this space as a certain degeneration of the K3 surface.}

In the light of the recently proposed AdS/CFT correspondence
\cite{adscft1,adscft2,adscft3} 
it is instructive to investigate domain walls in the 
Anti-de Sitter background. Sections~\ref{adssect} and
\ref{exoticadssect} are therefore devoted to the study of
four-manifolds of the abovementioned Bianchi types with negative
cosmological constant. 

\subsection{Bianchi models}

Let us now be more specific about the four-manifolds $M_4$ in question.
Spaces of interest are homogeneous manifolds with the following
ansatz for the metric:
\ben \label{bianchiansatz}
ds^2 = dt^2 + a^2(t)\,\left(\sigma^1\right)^2
+b^2(t)\,\left(\sigma^2\right)^2 +
c^2(t)\,\left(\sigma^3\right)^2\ . 
\een 
Here $t$ is the imaginary time and the metric coefficients are
functions of $t$ only. The one-forms $\{\sigma^k\}$, $k = 1,2,3$, are
left-invariant one-forms of the three-dimensional  group of isometric
motions $G$ and as such satisfy:
\[
d\sigma^k = -\frac{1}{2} \, n_k\, \epsilon_{ijk}\, \sigma^i\wedge
\sigma^j\ , \ \ {\rm no\ sum\ over\ }k\ ,
\]
where constants $\{n_k\}$  are the structure constants of $G$.
The four-manifolds may be classified according to the group of
isometric motions. This is the Bianchi classification in which each
type corresponds to a particular set of values of the structure
constants $\{n_k\}$. In the
rest of the paper manifolds of four
Bianchi types will arise, whose properties are  summarized in Table 
\ref{bianchitable}. 
\begin{table}
\begin{center}
\vskip 10pt
\begin{tabular}{|c|c|c|c|c|}
\hline
Bianchi type& $n_1$ & $n_2$ & $n_3$ & Group of motions \\
\hline
I& 0& 0& 0& \re$^3$ \\
II& 0& 0& 1& $Nil$ \\
VI$_0$& 1& $-1$ & 0 & $E(1,1)$ \\
VII$_0$ & 1& 1& 0& $E(2)$\\
\hline
\end{tabular}
\caption{Some of the Bianchi types} \label{bianchitable}
\vskip 10pt
\end{center}
\end{table}
 Note that all four groups of isometric motions
are solvable, in fact they all have one non-trivial commutator.
The \ein's equations for the metric (\ref{bianchiansatz}) reduce to
the  following set of second-order ODE's:
\begin{eqnarray}
-R^0_{\ 0}&  = & \frac{\ddot{a}}{a} +\frac{\ddot{b}}{b} +
\frac{\ddot{c}}{c}\ , \label{r00} \\
 R^1_{\ 1}&  = & \frac{(\dot{a} b c)^\cdot}{abc} + \frac{1}{2}
 \frac{1}{a^2 b^2 c^2}\, \left[n_1^2 a^4 - (n_2 b^2 - n_3 c^2)^2
\right]\ , \label{r11} \\
R^2_{\ 2}&  = & \frac{(a \dot{b} c)^\cdot}{abc} + \frac{1}{2}
 \frac{1}{a^2 b^2 c^2}\, \left[n_2^2 b^4 - (n_1 a^2 - n_3 c^2)^2
\right]\ , \label{r22} \\
R^3_{\ 3}&  = & \frac{(a b \dot{c})^\cdot}{abc} + \frac{1}{2}
 \frac{1}{a^2 b^2 c^2}\, \left[n_3^2 c^4 - (n_1 a^2 - n_2 b^2)^2
\right]\ , \label{r33}
\end{eqnarray}
where $\dot{a} = da/dt\ ,\ etc.$.
If the metric on $M_4$ is Ricci-flat, i.e. $R^a_{\ b} = 0$, equations
(\ref{r00}) - (\ref{r33}) are integrable in most cases. The
resulting manifolds are singular.\footnote{Many self-dual
  four-dimensional vacuum solutions of various Bianchi types have been
  found in \cite{lorentz}.} 

For non-Ricci-flat manifolds, in particular for manifolds with $R^a_{\
  b} =  \Lambda  \delta^a_{\ b}$, the \ein's equations are not in
  general integrable. However, a number of solutions with extra symmetries
exist. For example, in section \ref{adssect} we discuss the Bergmann
  metric ---  a
Bianchi type II  solution with negative cosmological constant. Unlike
the Ricci-flat Bianchi type II solution, the Bergmann metric is complete. 

\subsection{\ma\  equation}

From the theory of \ka\ manifolds it is known that the \ka\ metric may
be obtained from a real-valued function of complex holomorphic coordinates $z =
\{z^a\};\ \bar{z} = \{\bar{z}^a\}$ called the \ka\ potential:
\[
g_{a\bar{b}} = \partial_a\,\partial_{\bar{b}}\,K(z,\bar{z})\ .
\]
Here  $\partial_a = \partial/\partial z^a$ and $\partial_{\bar{b}}=\partial/\partial \bar{z}^b$.
A \ka\ manifold is \ein--\ka\ if the \ka\ metric $g_{a\bar{b}}$
satisfies the \ein's equations:
\[
{\cal R}_{a\bar{b}} = \Lambda\,g_{a\bar{b}}\ .
\]
These are equivalent to the requirement that the \ka\ potential
satisfy the so-called \ma\ equation obtained as follows.
The Ricci tensor is given by:
\[
{\cal R}_{a\bar{b}} =-\,\partial_a\,\partial_{\bar{b}}\log\,\det
g(z,\bar{z}))\ , 
\]
and hence the \ein--\ka\ condition reduces to
\ben \label{maeqn}
\det(\partial_a\,\partial_{\bar{b}}\,K) = e^{-\Lambda\,K}\ .
\een
In general this is a complex partial differential equation solving
which is not straightforward. If however the manifold possesses
certain amount of symmetry, the \ma\ equation may  reduce to an
ODE. In the following sections we shall deduce the \ma\ equations and
their solutions for most of the \ka\ manifolds that we study.

\section{Vacuum solutions of Bianchi type I and II} \label{vacuumsect}

 We shall begin by assuming
that the domain walls are invariant under three translations,
i.e. that they are Bianchi type I, or Kasner, but we shall find that
to be supersymmetric objects they should instead be invariant under
the nilpotent Bianchi type II 
group $Nil$ often called the Heisenberg group. 

\subsection {Kasner walls}

One's first idea might be to choose the metric
$g_{\alpha \beta}$ on $M_4$ to depend only on one ``transverse'' coordinate,
call it $t$, and to be independent of the other three coordinates $x^1,x^2,x^3$
say. Thus the metric would admit an isometric action of $\re^3$ or, if we
identify, $T^3$ and falls into the vacuum Bianchi I or Kasner class of
solutions \cite{kasner}: 
\ben \label{kasnermet} 
ds^2= dt^2 + t ^{2 \alpha_1} (dx^1)^2 + t ^{2 \alpha_2} (dx^2)^2+ t^{2
  \alpha_3} (dx^3)^2, 
\een
where constant $\alpha_1, \alpha_2, \alpha_3$ satisfy:
\ben \label{kasnerconstraint}
\alpha_1+\alpha_2+\alpha_3 = 1 = \alpha_1^2+\alpha_2^2 +\alpha_3^2\ .
\een
There are two problems with this metric. The first problem is that
if metric (\ref{kasnermet}) is  not flat, it is singular at the domain
wall $t=0$; and the second problem is that it is not BPS.

Consider, for example the rotationally symmetric case 
\[
(\alpha_1,\alpha_2,\alpha_3)= ({4 \over 3}, {4\over 3}, {2\over 3}) \ .
\]
Metric (\ref{kasnermet}) is then singular at $t=0$, but complete as
$t\rightarrow \infty$. Thus, in accordance with our
general remarks made in the Introduction, it is not asymptotically
flat in the usual sense 
although the curvature falls off as $t^{-2}$.

\subsection{BPS walls: Bianchi type II}  

To be BPS the manifold $M_4$ must admit at least one, and hence at least two,
 covariantly constant spinors.  If the solution admits
 at least one tri-holomorphic Killing vector it may be cast in the form:
\ben \label{harmonic}
ds^2= V^{-1} (d\tau + \omega_i dx^i)^2 + V d{\bf x}^2,
\een
where ${\bf x} = (x^1,x^2,x^3)$ with
\[
{\rm curl}\ \omega = {\rm grad}\ V.
\]
One may either   regard the ignorable coordinate $\tau$ as lying
in the world-volume of the $p$-brane or as a Klauza--Klein coordinate.
Obviously one may entertain both interpretations simultaneously in which 
 case one is considering the double-dimensional reduction
 of a brane in a lower dimensional space \cite{gibbonstownsend}. For
 the time being we will not tie ourselves
 down on this point. In order to get a domain wall solution
 we want some sort of invariance under two further translations
 and we are naturally led to choose for the harmonic function  
\[
V= z, 
\]
where we now interpret the coordinate $z$ as a  transverse coordinate.
With $x^1\equiv x,\,x^2\equiv y,\,x^3\equiv z$ metric (\ref{harmonic}) becomes:
\ben \label{heismet}
ds^2 = z dz^2 + z(dx^2 + dy^2) + z^{-1} (d \tau - xdy)^2 .
\een
The transverse proper distance is given by
\[
t= { 2\over 3}\ z^{3 \over 2}.
\]
The metric is complete as $z\rightarrow + \infty$,
the curvature again falls off as $t^{-2}$ but it  clearly has a
singularity at $z=0$,
at which the signature changes from $(++++)$ to $(----)$. We shall return
to this point shortly.

The \ma\ equation and the \ka\ potential for this metric will be given
in section~\ref{horospheresect} where suitable complex coordinates are
introduced.

\subsection {Geometrical considerations and the Heisenberg group}
\label{heissect} 

Evidently metric (\ref{heismet}) is not invariant
under translations in the $y$ coordinate; nevertheless it
admits a three-dimensional group of isometries. The metric may be written in
the general form (\ref{bianchiansatz}) so that the group of isometric
motions is manifest: 
\ben \label{heist}
dt^2+  \left({3t \over 2}\right) ^{-{2 \over 3}}   (\sigma^3)^2 + \left({3t
\over 2}\right) ^{{2 \over 3}} (  (\sigma ^1)^2 + (\sigma ^2)^2) \ ,   
\een 
where $\{\sigma ^k\}$ are left-invariant one forms on the $Nil$ or the
Heisenberg group. From this point on we shall refer to the metric
(\ref{heismet}) (or (\ref{heist})) as the {\em Heisenberg} metric. The
Heisenberg group may be defined as the 
nilpotent group $Nil=\{g\}$ of $3\times 3$ real-valued upper
triangular matrices:
\[
g= \pmatrix{ 1 & x & \tau \cr 0 & 1 & y \cr 0 & 0 & 1 \cr }. 
\] 
The Lie algebra of $Nil$ has as a basis\footnote{We adhere
  to the conventions that if a group $G$ with Lie algebra 
$\bigl [ { \bf  e }_a, {\bf e}_b \bigr ] = { {{C_a} ^c}}_b {\bf e}_c$
acts on the left on a
manifold $M$ then the   Killing vector fields  $ {\bf R}_a$ have Lie brackets 
$\big [{\bf R}_a, {\bf R}_b \big ]=-{{ C_a} ^c}_b {\bf R}_c$, while the
left-invariant one-forms $g^{-1} dg={\bf e}_a \, \sigma ^a$  satisfy
$d \sigma ^c =  -{1\over 2}{{ C_a} ^c}_b \sigma ^a \wedge \sigma ^b$.}: 
\begin{eqnarray}
{\bf e}_1& = & \pmatrix{ 0 & 1 & 0 \cr 0 & 0 & 0 \cr 0 & 0 & 0 \cr}
\nonumber \\
{\bf e}_2& = & \pmatrix{ 0 & 0 &  0   \cr 0 & 0 & 1  \cr 0 & 0 & 0 \cr}
\nonumber \\
{\bf e}_3 & = & \pmatrix{ 0 & 0 & 1  \cr 0 & 0 & 0  \cr 0 & 0 & 0 \cr}\nonumber
\end{eqnarray}
and the only non-vanishing commutator is 
\[
\Bigl [{\bf e}_1, {\bf e}_2 \Bigr ]= {\bf e}_3\ . 
\]  
The basis elements $\{{\bf e}_k\}$ correspond to three right-invariant
Killing vector fields 
\begin{eqnarray}
{\bf R}_1 & = & {\partial \over \partial x } + y {\partial \over
 \partial \tau} ,\nonumber \\ 
{\bf R}_2 & = & {\partial \over \partial y} \ ,\nonumber \\
{\bf R}_3 & = & { \partial \over \partial \tau}\ , \nonumber 
\end{eqnarray}
for which the only non-vanishing commutator is 
\ben 
\Bigl [ {\bf R}_1, {\bf R}_2 \Bigr ]= -\,{\bf R}_3 \ ;  
\een
and three left-invariant one-forms
\begin{eqnarray} \label{heis1forms}
\sigma ^1 & = & dx\ , \\
\sigma ^2 & = & dy\ ,\nonumber \\
\sigma ^3 & = & d\tau-xdy\ ,\nonumber
\end{eqnarray}
whence 
\[
d \sigma^3 = -\sigma ^1 \wedge \sigma ^2\ . 
\]
In addition this metric admits a rotational Killing vector of the form
\ben \label{rotvector}
{\bf m}= - x\frac{\partial}{\partial y} + y\frac{\partial}{\partial x} -
\frac{x^2 - y^2}{2}\frac{\partial}{\partial \tau} .
\een 
This Killing field $\bf m$ induces a rotation of $\sigma^1$ into
$\sigma^2$ but leaves $\sigma ^3  $ invariant.

The four-dimensional Heisenberg manifold (\ref{heismet}) is Ricci-flat
\ka\  and hence
carries a \hk\ structure. To exhibit the three \ka\ forms let us
introduce the following  orthonormal basis of one-forms $\{e^0, e^k\}$:
\begin{eqnarray} \label{frames}
e^0 & = & z^{-{1\over 2}}\,(d\tau - x dy)\ , \\ 
e^1 & = & z^{1\over 2}\, dx\ ,\nonumber \\
e^2 & = & z^{1\over 2}\, dy\ ,\nonumber \\
e^3 & = & z^{1\over 2}\, dz\ .\nonumber 
\end{eqnarray} 
In terms of these frames the three self-dual two-forms which are the
\ka\ forms are:
\ben \label{sdforms}
\Omega_x = e^0\, \wedge\, e^1 + e^2\, \wedge\, e^3\ , \ \ \ {\rm
  and\ cyclic\ permutations}\ ,
\een
and for the Heisenberg metric (\ref{heismet}) these become
\begin{eqnarray} \label{kahlerforms}
\Omega_x & = & (d\tau - xdy)\wedge dx + z\, dy\wedge dz\ ,\\
\Omega_y & = & (d\tau - xdy)\wedge dy + z\, dz\wedge dx\ , \nonumber \\
\Omega_z & = & (d\tau - xdy)\wedge dz + z\, dx\wedge dy\ . \nonumber
\end{eqnarray}
It is easily seen that the self-dual two-forms $\Omega_x,\, \Omega_y$
and $\Omega_z$ --- the \ka\ forms --- are closed and hence harmonic. 
They are clearly invariant under the action of  the Heisenberg 
group. However only  $\Omega_z$ is invariant under the circle action
generated by the rotational Killing field $\bf m$ (\ref{rotvector}).  

\subsection{Circle bundles and volume growth} \label{periods}

If one wishes to identify the coordinates $x$ and $y$ to 
obtain a two-dimensional torus one is forced to
make appropriate identifications of the coordinate $\tau$. The result is a
circle bundle over $T^2$. Such bundles $M_k$
 are indexed by an integer $k$
which is essentially the Chern class. They are often referred to as
{\em Nilmanifolds}.

If the periods of $(x,y,\tau)$ are $(L_x,L_y,L_\tau)$ then one must have: 
\ben
k= {L_x L_y \over L_\tau} \in {\Bbb Z}.  
\een 
If that is true then
\[
\exp(L_x {\bf e}_1) \exp(L_y {\bf e}_2) \exp(-L_x
{\bf e}_1)\exp(-L_y {\bf e}_2)=\exp (L_x L_y{\bf e}_3), 
\]
and hence  $\exp(L_x {\bf e}_1)$, $\exp(L_y {\bf e}_2)$ and $\exp(L_\tau
{\bf e}_3)$ will close on a discrete group ${\cal N}_k$. One  then has: 
\[
M_k= {Nil}/{\cal N}_k\ ,
\]
which clearly admits a global right action of $U(1)=\exp(z {\bf e}_3)$.

The curvature of the connection pulled back to the base $T^2$ is
\[
F=d\,\sigma ^3= dy \wedge dx
\]
and the Dirac quantization condition is
\ben \label{quant}
{ 1\over L_\tau} \int_{T^2} F=k \in \z\ .
\een
We shall see  later that the relevant value of  the integer $k$ in our
case is $k=3$. 
Formula (\ref{quant}) takes on a more conventional  appearance if one  chooses
$L_\tau=2\pi$. Alternatively, one could think of $e=2\pi/ L_\tau$
 as an electric charge.  It has been known for some time  that a
 Kaluza--Klein reduction on the Heisenberg group  gives
 rise to a uniform magnetic field \cite{nielsen}. Interestingly, since
 the present solution is BPS, it should be stable against production of
 monopole--anti-monopole pairs. This is in contrast to other examples
 of magnetic fields in Kaluza--Klein theories, for example in vacua
 studied in \cite{dowker93,dowker95} such
 monopole--anti-monopole pairs  are produced.

The curves of constant  $\{x, y, \tau\}$ are geodesics 
orthogonal to the group orbits and the coordinate $t$ is the radial distance.
 If the orbits  are compact
 we may  estimate how the four-volume of a geodesic ball increases with $t$
 by calculating the four-volume of the metric (\ref{heismet}) between
 $t=t_1$  say and $t$. This is easily seen to grow with $t$ as $t^{4 \over 3}$.
 We shall use this fact in section~\ref{blowupsect} to compare with 
 the work of Bando, Kobayashi, Tian and Yau
 \cite{tianyau89,tianyau90,bandokobayashi88,bandokobayashi90} on an exact
 metric on the complement of a smooth cubic curve in $\co \p^2$.

\section{Resolution of the singularity}
In this section we describe the  physical motivation for resolving the
singularity of the Heisenberg manifold
(\ref{heismet}) and analyse the underlying mathematical structure of
the proposed resolution.

\subsection{8-branes, 6-branes and T-duality}

The metric (\ref{heismet}) 
has been  reached previously by a different route. The massive type
IIA  ten-dimensional
theory of Romans \cite{romans} admits BPS solutions corresponding to
Dirichlet  8-branes whose properties  
have been discussed by Polchinski and Witten \cite{polchinskiwitten}
and Bergshoeff {\em et al} \cite{bergshoeff}. The
solutions are based on a harmonic
function  of the coordinate transverse to the 8-branes which has
discontinuities at the location of the 8-branes. The relation of Romans'
theory to eleven-dimensional supergravity theory is
unclear.\footnote{See a very recent paper of Hull \cite{hull98}.}
However under
a double T-duality with respect to two coordinates lying in the 8-brane,
$x$ and $y$ say, it may be reduced to a 6-brane solution of the IIA
theory compactified to eight dimensions. Under T-duality,
the coordinates $x$ and $y$  become transverse coordinates and,  strictly
speaking, because the solution is independent
 of the coordinates  $x$ and $y $  one  has a superposition of 6-branes.
A 6-brane solution of the ten-dimensional type IIA theory 
 may be lifted to eleven dimensions to give
 a BPS  7-brane wrapped around  the eleventh dimension. In other
 words the eleven-dimensional 7-brane is
 a Ricci-flat metric of the form: 
\ben
\euc ^{\, 6,1} \times M_4\ ,
\een
where $M_4$ is a multi-Taub--NUT metric of the form (\ref{harmonic}):
\[ 
ds^2= V^{-1} (d\tau + \omega_i dx^i)^2 + V d{\bf x}^2,
\]
with
\[
{\rm curl}\ \omega = {\rm grad}\ V\ .
\]
The coordinate $\tau$ is the eleventh direction.
Coordinates ${\bf x}$ are transverse to the 6-brane.
 A single 6-brane corresponds to the Taub--NUT metric with positive mass which
has 
\[
V=1+ { 1\over r}\ .
\]
In order to get a superposition of 6-branes which is independent
of $x$ and $y$  (at least up to gauge transformations) one should choose 
\[ 
V=z\ , 
\]
and this is indeed what Bergshoeff {\em et al} \cite{bergshoeff} find.

\subsection{Sources} 

As it stands, metric (\ref{heismet}) is singular at $z=0$. In fact
this singularity resembles the singularity in the self-dual
Taub--NUT metric with negative mass parameter  for which $V= 1-1/ r$
in (\ref{harmonic}).
 On the three-surface $r=1$ the metric  changes signature from $(++++)$ to
 $(----)$. The Taub--NUT metric with negative mass parameter is known
 to be  asymptotic
 to a complete topologically non-singular self-dual
 Riemannian manifold called the
Atiyah--Hitchin  manifold \cite{atiyahhitchin}. The presence of the
singularity at
$r=1$ is a clear indication of the fact that the Taub--NUT approximation is
broken already at values of $r$ greater than one.  It is natural to
suppose that something similar may be happening in the case of the
Heisenberg metric (\ref{heismet}). 
 Indeed in the next section we
shall make a concrete proposal  for the exact metric.
However Bergshoeff {\em et al} \cite{bergshoeff} and others
writing on supergravity domain walls \cite{domainwalls} do something else. They
replace $z$ by $|z|$ which results in a configuration 
symmetric under the reflection $z \rightarrow -z$.
 The justification for this procedure
 is that one has inserted a distributional source at $z=0$
 representing the domain wall and the regions $z >0$ and $z< 0$ 
 correspond to the two sides of the domain wall.  Geometrically this
resembles but is not equivalent to the procedure of Israel \cite{israel}
used in classical general relativity  who describes a shell of matter by
gluing together two smooth spacetimes $M^\pm$
across a hypersurface $\Sigma$. The Israel matching conditions are that
the two metrics $g^\pm _{ij}$ induced on $ \Sigma$  from $M^\pm$ agree. One
then evaluates the distributional stress tensor from the discontinuity in the
second fundamental forms $(K^+_{ij}-K^-_{ij})$ across $\Sigma$.
From the point of view of M-theory there are two objections to doing this in
the present case:
\begin{itemize}
\item There are no obvious sources in M-theory
\item The induced metric on the hypersurface $\Sigma$ given by $z=0$
  is  singular.
\end{itemize}

\subsection {Orbifold walls}

An alternative attitude to the singularity of (\ref{heismet}) at $z=0$
would be to identify the region $z > 0$ with the region $z < 0$.
The singularity would then be viewed as a consequence the fact that the
reflection has a fixed point set. Thus one has something analogous to the two
orbifold domain walls at the ends of an interval in Ho\v{r}ava and Witten's
compactification of the eleven-dimensional $M$-theory  on $S^1\times \z_2$ to
give the $E_8 \times E_8$ heterotic theory in ten dimensions
\cite{horavawitten1,horavawitten2}. 
In the formulation of \cite{ovrut} one considers the
eleven-dimensional metric on $\euc^{3,1}\times S^1/\z^2\times X^6$
\[
ds^2 = \frac{1}{H}\, g^4_{\mu\nu}+ H^2\,dy^2 + H\, g^6_{AB}\ ,
\]
where $g^4_{\mu\nu}$ is the four-metric on the flat Minkowski space-time
$\euc^{3,1}$, $y$ is the coordinate on the interval $S^1/\z^2$ ranging
from $-\pi \rho$ to $\pi \rho$, and $g^6_{AB}$ is the metric on the
compact Calabi--Yau space $X^6$. Function $H$ is a harmonic function
linear in $y$ and invariant under the reflection $y\rightarrow
-y$. In addition, there is a non-vanishing four-form field strength in
the eleven-dimensional theory. In the effective five-dimensional
theory obtained by generalized Kaluza--Klein dimensional reduction on
the Calabi--Yau space $X^6$, this solution can be viewed as a pair of
3-brane domain walls on the orbifold fixed planes $y = 0$ and
$y=\pi \rho$. The 3-branes are in fact the M-theory 5-branes
with two world-volume dimensions ``wrapped'' on a two-cycle in $X^6$. 

In our case the solution is defined on $ \euc^{\,6,1}\times
\re_{+}\times Nil$, where $\re_{+}$ is parametrized by $z>0$ and the
three-manifold $Nil$ parametrized by $\{x,y,\tau\}$ is the group manifold
of the Heisenberg group. We may think of this as a 9-brane solution
of eleven-dimensional supergravity where the world-volume of the
9-brane is taken to be $Nil\times \euc^{6,1}$. Replacing $Nil$ by
$M_k = Nil/{\cal N}_k$ defined in section~\ref{periods} amounts to
``wrapping'' the 9-brane on the $S^1$ bundle over $T^2$. Just as in
the Ho\v{r}ava--Witten case we do not have the full $SO(9,1)$ Lorentz
invariance, rather it is broken to $SO(6,1)\times Nil$.  

\subsection{Scherk--Schwarz reduction to seven dimensions}

In the light of the comments above, particularly the absence of
Lorentz invariance, perhaps the most attractive interpretation of the
solution (\ref{heismet}) is that adopted by Lavrinenko {\em et al}
\cite{pope97}.
One regards it aa a  solution of the 
so-called ``massive'' eight-dimensional theory that is obtained by reducing
 eleven-dimensional  supergravity  {\it \'a la}
 Scherk and  Schwarz \cite{scherkschwarz}. In other words, one restricts
 the  eleven-dimensional
 theory to solutions invariant under the action of the three-dimensional
 Heisenberg group. The resulting theory has a potential
 for the  scalar fields arising from the reduction
 and as a consequence there is no solution with the eight-dimensional
 Poincar\'e invariance. Lavrinenko {\em et al} \cite{pope97} therefore
 propose using the BPS
 solution (\ref{heismet}). In their interpretation $z$ is,
 as with us, the transverse coordinate (i.e. the eighth coordinate) and 
 $\{x,y,\tau\}$ are the ninth, tenth and eleventh coordinates
 in no particular order. Since the size of the $x$ and $y$ directions
 goes to infinity as $z\rightarrow \infty$ and the size of the $\tau$
 direction goes to zero, the Scherk--Schwarz reduction is not really a
compactification even if one identifies the coordinates so as to
obtain a circle bundle over a two-torus. 
 It is however certainly a consistent truncation of the theory.

\subsection {Blowing up the singularity} \label{blowupsect}

If the configuration (\ref{heismet}) really does come from M-theory
we still face the problem of the source. We have two possibilities:
\begin{itemize}
\item {\sl Either} to follow Bergshoeff {\em et al} \cite{bergshoeff} and
  Lavrinenko {\em et al} \cite{pope97}  and take the view
that the domain wall has two sides. 
\item {\sl Or} to adopt the  orbifold interpretation and
  identify the regions of positive and negative  $z$. 
\end{itemize}

Both approaches give rise to a singularity. The question arises as to
whether one can  somehow smooth
out the singularity? 
We are going to argue that the answer is {\em no} if we adopt 
 the first course and {\em yes} if we adopt the second.
Assuming that only gravity with no extra form-fields is present,
we thus seek a non-singular Ricci-flat
BPS metric which is asymptotic to the Heisenberg metric
(\ref{heismet}).
 
 To see that the first approach is ruled out we note that if the
 singularity could be resolved, then keeping coordinates
 $\{x,y,\tau\}$  would give a 
 complete Ricci-flat metric on $\re \times \Sigma$ where
 $\Sigma $ is a closed complete three-manifold. In particular, the manifold
 would have two 
``ends'', i.e. two infinite regions. However if this were true we could
 construct a  ``line'' between the two ends, that is a geodesic which
 minimizes the length between any two points lying on it. But by the
 Cheeger--Gromoll Theorem this is impossible (see
 e.g. \cite{besse}). Thus we are
 forced to adopt the second course of action which is 
investigated in detail in the following section. 

Before doing so it is perhaps worthwhile pointing out the analogy
of the situation in question with the case of the blow up of $\euc^4 / \z_2$.  
One might have wondered if it is possible to glue together two
copies of $\euc^4$ to get a Ricci-flat
 wormhole-like structure with topology $\re \times \re \p^3$. 
Again by the Cheeger--Gromoll theorem  this cannot happen. In fact
 we know that the correct blow up of $\euc^4 / \z_2$ is the
 Eguchi--Hanson manifold
 on the cotangent bundle of $\co \p^1$ \cite{eguchihanson}
 and that this manifold has only one infinite region.

 Another completely analogous situation is the  
 Taub--NUT approximation to an orientifold plane. This is obtained by
 taking the metric (\ref{harmonic}) with $V = 1-1/r$ and making a
 further identification \cite{sen}. The metric is incomplete because
 of the singularity at $r=1$. This singularity cannot be resolved by 
 joining together two  copies of the Taub--NUT metric
across $r=1$ because this would also produce a manifold with  two ends.
 The correct way to blow up the singularity of the Taub--NUT metric is
 to pass to the Atiyah--Hitchin manifold.

\subsection{Complement of a cubic in $\co \p^2$: the BKTY metric}
 \label{bktysect} 

We now turn to the problem of finding, or more
 properly speaking identifying, the exact metric of which the Heisenberg
 metric (\ref{heismet}) is an asymptotic approximation. This task is
 greatly facilitated by the extremely helpful review
 of Kobayashi \cite{kobayashi} on degenerations of the metric on
 K3 and in what follows we shall rely heavily on that reference.
The general self-dual four-metric on K3 has (including an overall scale) a 58
parameter moduli space. As we move to the boundary of the moduli space in
certain directions the four-metric may decompactify, 
while remaining complete and non-singular. Among the degenerations discussed
by Kobayashi there is one he refers to as  type II.
It may be constructed by considering the complement
$M_4= \co \p^2\setminus C$  of a smooth cubic curve $C$  in the
complex plane $\co \p^2$. This has a \ka\ metric: the
Fubini--Study metric which is incomplete because the cubic has been removed.
However using general existence theorems 
for solutions of the \ma\ equation
 Yau, Tian, Bando and Kobayashi \cite{tianyau89}--\cite{bandokobayashi90}
 have shown that there exists a complete non-singular Ricci-flat \ka\
 (and hence self-dual) metric on $M_4$. Clearly the metric must blow
 up on the cubic $C$ which corresponds to infinity.
 
Consider now the neighbourhood of the cubic $C$. The curve itself is
topologically a two-torus $T^2$. A normal neighbourhood
consists of a  disc bundle over $T^2$. The centre of the disc
corresponds to infinity in $M_4$. The radial direction corresponds to
a geodesic in the  self-dual metric. A surface of constant radius
is a circle bundle over the torus. This is the three-dimensional Nilmanifold.  
 
 Kobayashi tells us that the the Nilmanifold collapses 
 as  we approach infinity in such a way that the metric spheres are an
 $S^1$-bundle over $T^2$, 
the size of the $S^1$ falls off as $t^{-1/3}$ ($t$ is the radial
 distance) and the size of each
cycle in $T^2$ grows as $t^{1/3}$. The volume of a metric ball
grows as $t^{4/3}$. This is exactly the behaviour of 
 the Heisenberg  metric (\ref{heist}) . It is therefore very plausible
 that the metrics constructed by Yau, Tian, Bando and Kobayashi
 do indeed asymptote to the metric (\ref{heist}). In what follows we shall
 assume that this is true. 
 
 The topology of $M_4$ is non-trivial\footnote{We thank Ryushi Goto
 for this computation.}: it is not simply connected and has 
\[ 
H_1(M_4) = \z_3 \quad \quad H_2(M_4) =\z \oplus \z.
\]
 Hence if  arguments
like those in \cite{hawkingpope} apply,
the manifold should admit at least two normalizable anti-self-dual
two-forms. Using the analysis of  \cite{hawkingpope} one deduces that
there should be a $2\times 3 =6$ dimensional family of  transverse traceless
zero modes of the Lichnerowicz operator. Adding the trivial overall
scaling we expect to find a seven-dimensional family of metrics. 

\subsection{Gravitational action} \label{actionsect}

Complete Ricci-flat (vacuum) \ein\ manifolds are gravitational
instantons. It is of interest to estimate their contribution to the
path integral of the Euclidean Quantum Gravity by evaluating their
gravitational action.  If $\cal M$ is a  non-compact manifold or a
compact manifold with boundary $\partial {\cal M}$ the gravitational
action is:
\ben \label{gravaction}
-\frac{1}{16\pi} \int_{\cal M} R - \frac{1}{8\pi}\int_{\partial {\cal
    M}} Tr {\cal K}\ , 
\een
where $R$ is the Ricci tensor  and $\cal K$ is the second fundamental form
on $\cal M$. The first term is the contribution from the bulk which
vanishes for Ricci-flat manifolds; the second term is the
contribution from the boundary (possibly boundary at infinity).
Traditionally only four-manifolds were
regarded as gravitational instantons, however expression
(\ref{gravaction}) is valid
for complete Ricci-flat  Einstein manifolds in any dimension.

Let us estimate the contribution from the boundary. Let $\bf n$ be a vector
normal to the boundary $\partial {\cal M} $, then the second term in
(\ref{gravaction}) is:
\[
\frac{1}{8\pi}\int_{\partial {\cal M}} Tr {\cal K} =
\frac{1}{8\pi}\frac{\partial}{\partial {\bf n}}\, ({\rm Vol}\, \partial{\cal M})\ . 
\]
By ${\rm Vol}\,\partial {\cal M}$ we mean the unit volume of the boundary.
For four-dimensional Ricci-flat manifolds, if $t$ is the
radial distance  the boundary term contribution  
to the action is finite if ${\rm Vol}\,\partial{\cal M}$ be no faster
than linear in $t$. This implies that the volume growth of a large
metric sphere at infinity should grow no faster than $t^2$. 
Similarly, for higher-dimensional instantons the ``critical'' volume
growth for which the boundary contribution to the action is finite
(but not necessarily vanishing) is $t^2$.

The BKTY manifold possesses a
 non-compact complete Ricci-flat \ka\ metric with the Heisenberg end
 and  can thus be viewed as a gravitational instanton.
Since it is Ricci-flat, its gravitational action receives no
contribution from the first term in (\ref{gravaction}).
At  infinity the BKTY metric looks like the Heisenberg metric (\ref{heist}). 
The boundary of (\ref{heist}) at large values of $t$ looks like an $S^1$
bundle over $T^2$, where the  two-torus is parametrized by $x$ and
$y$; $\tau$ is the fibre coordinate. Hence the second term in
(\ref{gravaction}) is:
\[
\frac{1}{8\pi}\int_{\partial {\cal M}} Tr {\cal K} =
\frac{1}{8\pi}\frac{\partial}{\partial {\bf n}}\, ({\rm Vol}\,
\partial{\cal M}) = 
\frac{1}{8\pi}\frac{\partial}{\partial\, t}\, \left[\left(\frac{3}{2}\,
t\right)^{1/3}\, {\cal V}\right]\ . 
\]
${\cal V} = L_x\,L_y\,L_\tau$ where $L_x, L_y$ and $L_\tau$ are the periods of
$x,y$ and $\tau$ as described in section \ref{periods} and we get
\[
\left(\frac{3}{2}\,t\right)^{-{2\over 3}}\, \frac{L_x L_y
  L_\tau}{16\pi}\ .
\]   
Note that the boundary unit volume grows as $t^{1/3}$ which is slower
than the critical estimate $t$, hence it is not surprising that
the gravitational action of the
BKTY instanton is finite and tends to zero as $t\rightarrow +\infty$.

\section{Bianchi types I and II with negative cosmological constant}
\label{adssect}

In this section we look for $p$-brane solutions of the form $M_4\times
\euc^{\, p-3,1}$, where now $M_4$ is not a Ricci-flat manifold but
rather a four-manifold with negative cosmological constant.
Such solutions may be interpreted as branes in the Anti-de Sitter
background and are likely to be of interest in connection with the
AdS/CFT \cite{adscft1,adscft2,adscft3} correspondence.

Here we focus on four-manifolds of Bianchi types I and II. The
relevant metrics must be solutions of the Einstein's equations
(\ref{r00})-(\ref{r33}) with $R^a_{\ b} = \Lambda \delta^a_{\ b}$,
$\Lambda<0$, and
appropriate values of the structure constants $n_k$ given in Table
\ref{bianchitable}. 

\subsection{Bianchi type I} \label{kasnerads}

In this case $n_k = 0$ and the \ein's equations are integrable. While there
is no polynomial solutions like the Kasner metric (\ref{kasnermet}),
the solution is obtained by replacing $t^{\alpha_k}$ in
(\ref{kasnermet}) with
\[
\left(\frac{\sinh(\sqrt{-3\Lambda}\,t)}{\tanh\left(\frac{\sqrt{-3\Lambda}}{2}\,
    t\right)}\right)^{1\over 3}\, \left(\tanh\left(\frac{\sqrt{-3\Lambda}}{2}\,
    t\right) \right)^{\alpha_k}\ .
\] 
where the powers $\alpha_k$ again satisfy equation (\ref{kasnerconstraint}). 
Setting $\alpha_1 = 1$, we get a complete non-singular (in contrast
with the singular Kasner metric) instanton if the
coordinate $x$ is suitably identified. This Kasner--Anti-de Sitter
metric could be used to construct domain walls. Note, however, that
like its vacuum counterpart (\ref{kasnermet}) this metric is not BPS.

\subsection{Bianchi type II: the Bergmann metric} \label{bergmannsect}

Substituting the relevant structure constants into the \ein's
equations (\ref{r00})-(\ref{r33}) we find 
$b(t) = c_0\, a(t) + c_1$. However we are only interested in self-dual
metrics (self-dula four-metrics are \hk) and for these the constants
$c_0, c_1$  are $c_0 = 1, c_1 =
0$. Hence we necessarily have $b(t) = a(t)$ and the \ein's
equations reduce to:
\begin{eqnarray} 
-\,\Lambda & = & 2\, \frac{\ddot{a}}{a} + \frac{\ddot{c}}{c}\ ,
\label{berg0} \\
-\,\Lambda & = & \frac{(\dot{a}a c)^\cdot}{a^2 c} - \frac{c^2}{2 a^4}\ ,
\label{berga} \\
-\,\Lambda & = & \frac{(\dot{c}a^2)^\cdot}{a^2 c} + \frac{c^2}{2 a^4}\ .
\label{bergc} 
\end{eqnarray}
It is not straightforward to solve equations
(\ref{berg0})-(\ref{bergc}), in fact it is not clear whether they are
integrable in general. There exists however a special solution for
which 
\[
a(t) = A\,e^{\alpha t}\ , \ \ \ c(t) = B\,e^{\gamma t}\ ,
\]
where $\alpha,\gamma,A,B$ are constants. Substituting this ansatz into
(\ref{berg0})-(\ref{bergc}) we find:
\[
\alpha^2 = -\,\frac{\Lambda}{6} = \frac{B^2}{4A^4}  , \ \ \ \gamma = 2\,\alpha .
\]
Since only the ratio $A/B$ plays a role we are free to choose $A = 1$,
giving $B = \sqrt{-\,\frac{2\Lambda}{3}}$. 
Clearly, the solution is invariant under time-reversal $t\rightarrow
-t$; it therefore 
suffices to consider negative parameters $\alpha <0$. The Bianchi type II
four-metric thus becomes:
\ben \label{Lbergmet}
ds^2 = dt^2 + e^{-2\,t\sqrt{-\,\frac{\Lambda}{6}}}\, \left((\sigma^1)^2 +
  (\sigma^2)^2\right) + \sqrt{-\,\frac{2\Lambda}{3}}
e^{-4\,t\sqrt{-\,\frac{\Lambda}{6}}}\,(\sigma^3)^2\ , 
\een   
where $\{\sigma^k\}$ are the left-invariant one-forms
(\ref{heis1forms}). A convenient choice of the cosmological constant
is $\Lambda =- 2/3$, for which (\ref{Lbergmet}) becomes:
\ben \label{bergmet} 
ds^2 = dt^2 + e^{-t}\,\left((\sigma^1)^2 + (\sigma^2)^2\right) +
e^{-2t}\,(\sigma^3)^2\ . 
\een
In (\ref{bergmet}) one recognises the Bergmann metric --- a
non-compact \ka\ symmetric space $SU(2,1)/U(2)$. Note that the
Bergmann metric is complete (unlike the vacuum Bianchi type II metric
(\ref{heismet})). It is also BPS.

Another form of the Bergmann metric where the $U(2)$ group action is
manifest is:
\ben \label{u2bergmann}
ds^2 = \frac{dR^2}{(1+\frac{\Lambda}{6}\,R^2)^2} +
\frac{R^2}{4\,(1+\frac{\Lambda}{6}\,R^2)^2}\, (\sigma^3)^2 +
\frac{R^2}{4\,(1+\frac{\Lambda}{6}\,R^2)} \, \left((\sigma^1)^2
  +(\sigma^2)^2 \right)\ .
\een 
Expressed in this form the metric is a limiting case of the general
$U(2)$-invariant Taub--NUT--Anti-de Sitter family of metrics
\cite{chamblin}  when ${\cal N}\rightarrow \infty$ (${\cal N}$ is the
NUT charge).  
Such a family can be written in the form (see equation (2.6) in reference
\cite{chamblin}):
\ben \label{tnadsmet}
\frac{U(r)}{f(r)}\, dr^2 + 4n^2\,\frac{f(r)}{U(r)}\,(\sigma^3)^2 +
r^2\,U(r)\,\left((\sigma^1)^2 +(\sigma^2)^2 \right)\ , 
\een
where $f(r) = 1+ \frac{\Lambda}{3}\,r^2\,(1 + \frac{4{\cal N}}{r})$
and $U(r) = 1 + \frac{2{\cal N}}{r}$. Now, to take the limit of large
NUT  charge  rescale the
radial coordinate $r = \rho/{\cal N}$ in the above formulae and take
${\cal N}\rightarrow \infty$. Metric (\ref{tnadsmet}) becomes:
\ben \label{rbergmann}
ds^2 = \frac{4\, d\rho^2}{2\rho\, \left(1+ \frac{4\Lambda}{3}\,\rho\right)} +
2\rho\, \left(1+ \frac{4\Lambda}{3}\,\rho\right)\,(\sigma^3)^2 +
2\rho\,\left((\sigma^1)^2 +(\sigma^2)^2 \right)\ ,  
\een
and taking $2\rho  = \frac{R^2}{4\,(1+\frac{\Lambda}{6}\,R^2)}$ we get
back to expression (\ref{u2bergmann}). 

\subsection{Horospheres}  \label{horospheresect}

To elucidate the geometrical structure of the Bergmann manifold and
the role played by the Heisenberg group we shall now describe the way
the Bergmann manifold arises as the set of horospheres of an
odd-dimensional Anti--de Sitter space. 
We shall also obtain the \ma\ equation for a Bianchi type II
four-manifold. A solution to the equation with a negative cosmological
constant gives the \ka\ potential for the Bergmann metric. We make use
of the defined complex coordinates and solve the Ricci-flat \ma\
equation to obtain the \ka\ potential for the Heisenberg metric
(\ref{heist}). 

Suppose $G/H$ is a non-compact Riemannian symmetric space. The Iwasawa
theorem (see e.g. \cite{barutraczka}) tells us that every element
$g\in G$ may be uniquely expressed as 
\[
g = h\, a\, n\ ,
\]
where $h\in H\subset G$, $a\in A\subset G$, $n\in N\subset G$; $\ A$
is abelian and $N$ is nilpotent subgroups of $G$. Moreover we have 
\[
G = H \times (A\ltimes N)
\]
where $\ltimes$ denotes a semi-direct product and  $A\ltimes N$ is a
solvable group. This means that we may
also think of $G/H$
as a solvable group $G_{solv} = A\ltimes N$ with a left-invariant
metric. The orbits of $N$ in $G/H$ are called {\em horospheres}. The set of
horospheres is labelled by elements of $A$. They are permuted by
elements of $H$. The simplest example would be an $n$-dimensional
real hyperbolic space $G/H = {\cal H}^n$, $A =
\re_{+}$ and $N = \re^{n-1}$. If we think of ${\cal H}^n$ as a quadric in
$\euc^{\, n,1}$ ($(n+1)$-dimensional Minkowski space-time) then
the horospheres are the intersections of the quadric with a family of
parallel null hypersurfaces related by boosts. There is a similar
description for an $n$-dimensional Anti--de Sitter space AdS$_n$
regarded as a quadric in  $\euc^{\, n-1,2}$. This description of the
hyperbolic and the Anti--de Sitter spaces will be useful in
section~\ref{bianchiIIIsect}.

The case of a complex hyperbolic $n$-space ${\cal H}^n_{\co}$ is slightly
more complicated. Thinking of $\euc^{\,2n,2}$ as $\co^{\, n,1}$,
$(2n+1)$-dimensional Anti--de Sitter space, AdS$_{2n+1}$, is given by
the quadric
\[
|z^0|^2 - \sum_{a = 1}^n |z^a|^2 = 1\ .
\]
Then the complex hyperbolic $n$-space ${\cal H}^n_{\co}$ is obtained by
identifying $z^a$ with $e^{i\theta}\, z^a$, $a = 1,\ldots,n$.
Thus $z^0,\ldots,z^n$ are homogeneous coordinates on ${\cal H}^n_{\co}$.
The nilpotent group $N$ turns out to be the
Heisenberg group. Let us see how this works
in detail. It is helpful to recall that the
inhomogeneous coordinates $\zeta^a$ are defined in the usual way as
$\zeta^a = z^a/z^0$ and make manifest the action of $U(n)$ on ${\cal
  H}^n_{\co}$.  Our aim is to find a set of coordinates to make the action
of the Heisenberg group $N$ manifest.  

Let us first introduce complex null coordinates $u$ and $v$
\begin{eqnarray}
u & = & \frac{1}{\sqrt{2}}(z^0 + z^n)\ , \nonumber\\
v & = & \frac{1}{\sqrt{2}}(z^0 - z^n)\ . \nonumber
\end{eqnarray}
Define $z$ and $w^i$, $i = 1,\ldots, n-1$  to be
\[
z = \frac{u}{v}\ ,\ \ \ w^i = \frac{z^i}{v}\ .
\]
In terms of the inhomogeneous coordinates $\{\zeta^i, \zeta^n\}$ these are
\begin{eqnarray} \label{heiscoords}
z & = & \frac{2}{1-\zeta^n} - 1 \ , \\
w^i & = & \frac{\sqrt{2}\, \zeta^i}{1-\zeta^n}\ . \nonumber
\end{eqnarray}

A complex hyperbolic $n$-space is topologically the interior of
a  unit ball in $\co^n$ and the map $(\zeta^i, \zeta^n) \rightarrow
(w^i,z)$ provides a  bi-holomorphism from the interior of the unit
ball in  $\co^n$ into the interior of the paraboloid
\[
z+\bar{z} > \sum_i\,|w^i|^2\ .
\]
As we have mentioned above, AdS$_{2n+1}$ is obtained from ${\cal
  H}^n_{\co}$ as the base of the Hopf fibration, with the time-like Hopf fibre
  parametrized by $\theta$ such that
\begin{eqnarray}
u & = & \frac{z^0 + z^n}{\sqrt{2}} = \frac{z\, e^{i\theta}}{(z+\bar{z} -
  \sum_i\,|w^i|^2)^{1\over 2}}\ , \nonumber\\ 
v &  = &\frac{z^0 - z^n}{\sqrt{2}} = \frac{e^{i\theta}}{(z+\bar{z} -
  \sum_i\,|w^i|^2)^{1\over 2}}\ , \nonumber \\
z^i & = & \frac{w^i\, e^{i\theta}}{(z+\bar{z} - \sum_i\,|w^i|^2)^{1\over 2}}\ .
\end{eqnarray}
The real quantity $(z+\bar{z} - \sum_i\,|w^i|^2)$ is invariant under
the action of the Heisenberg group $N$ parametrized by $(a^i,b)$:
\begin{eqnarray} \label{Naction}
w^i & \rightarrow & w^i + a^i\ , \\
z & \rightarrow & z + i\, b + \sum_i {1\over 2} |a^i|^2 + \bar{a}^i\,
w^i\ . \nonumber
\end{eqnarray}
Considered as a subgroup of $SU(n,1) \subset SO(2n,2)$ $\ N$ acts on ${\cal
  H}^n_{\co}$ as 
\begin{eqnarray}
u & \rightarrow & u + (i\, b + \sum_i {1\over 2} |a^i|^2)\, v + \bar{a}^i\,
z^i \nonumber\\
v &\rightarrow & v \nonumber\\
z^i & \rightarrow & z^i + a^i\ . \nonumber 
\end{eqnarray}
Finally, the abelian group $A = \re_{+}$ parametrized by $\lambda$
acts as $(z, w^i) \rightarrow (\lambda^2\,z, \lambda\, w^i)$ or
\begin{eqnarray}
u & \rightarrow & \lambda\, u\ , \nonumber\\
v &\rightarrow & \lambda^{-1}\, v\ , \nonumber\\
z^i & \rightarrow & z^i\ . \nonumber 
\end{eqnarray}
Having identified all group actions clearly, let us now formulate
the problem in terms of the \ka\ potential.
The \ka\ potential for the metric on the horospheres may be obtained
from the \ka\ potential on the AdS$_{2n+1}$ manifold which in terms of
inhomogeneous coordinates $\zeta^a$ is given by 
\[
K(\zeta^a, \bar{\zeta}^a) = -\log(\sum_{i=1}^{n-1} |\zeta^i|^2
+|\zeta^n|^2 - 1)\ . 
\]
From (\ref{heiscoords})
\[
\sum_i |\zeta^i|^2 +|\zeta^n|^2 - 1 = 2\,\frac{|w^i|^2 -
  (z+\bar{z})}{(z+1)(\bar{z}+1)}  
\]
and hence the \ka\ potential on the horospheres becomes (up to a \ka\ gauge
transformation)
\ben \label{bergpot}
K = -\log(z+\bar{z} - \sum_i\,|w^i|^2)\ . 
\een

Let us derive the \ma\ equation to which \ka\ potential
(\ref{bergpot}) is a solution.
Since the resulting
metric  contains the higher-dimensional extension of the 
Heisenberg group (see section~\ref{highdimsect}) as
a group of isometries we must 
assume that the \ka\ potential depends only on the real quantity $f =
z+\bar{z} - \sum_i\,|w^i|^2$, the \ma\  equation (\ref{maeqn}) becomes
an ordinary differential equation: 
\ben \label{maeqnIIhigh}
(K')^{n-1}\,K'' = (-1)^{n-1}\,e^{-\,\Lambda\,K}\ ,
\een
where $K' = dK/df$.

Let us make the 
connection with the form of the four-dimensional Bergmann metric
(\ref{bergmet}). In this case $n=2$ and there are two complex
coordinates $(z,w)$. We can pass from this parametrization to the
parametrization of (\ref{bergmet}) in terms of $\{t,x,y,\tau\}$ as follows:
\begin{eqnarray} \label{bergmap}
z - \bar{z} & = & i\left(\tau - \frac{x\,y}{2} \right)\ , \\
z + \bar{z} & - & w\bar{w} =  f\ ,\nonumber \\
w & = & {1\over 2} (x+iy)\ ,\nonumber \\
t & = & e^f\ .\nonumber 
\end{eqnarray}
For $n=2$ the \ma\ equation (\ref{maeqnIIhigh}) becomes
\ben \label{maeqnII}
K'\,K'' = -\,e^{-\,\Lambda\,K}\ .
\een 
In terms of the \ka\ potential $K(f)$ the compatible \ka\ metric is
\ben \label{metricII}
ds^2 = - K'\,dw\wedge d\bar{w} + K''\,(dz - \bar{w}\,dw)(d\bar{z} -
  w\,d\bar{w})\ .
\een
The \ka\ potential (\ref{bergpot}) for $n=2$ 
\ben \label{kpotberg}
K = -\log f = -\log\log t
\een
is clearly a solution of equation (\ref{maeqnII}) for $\Lambda  = -3$.  
The \ka\ metric $g_{a\bar{b}} = \partial_a\, \partial_{\bar{b}} K$---
the Bergmann metric in complex coordinates ---
obtained from the \ka\ potential (\ref{kpotberg}) is:
\[
ds^2 = \frac{1}{f}\,dw\wedge d\bar{w} + \frac{1}{f^2}\,(dz -
  \bar{w}\,dw)(d\bar{z} -  w\,d\bar{w})\ .
\]
Rewriting this metric using definitions (\ref{bergmap})
we get  the Bergmann metric in the standard form (\ref{bergmet}).

Let us solve the \ma\ equation (\ref{maeqnII}) in the Ricci-flat
case $\Lambda = 0$. Here we present the four-dimensional case $n=2$ leaving
the treatment of the higher-dimensional example to
section~\ref{highdimsect}. The solution of (\ref{maeqnII})should yield
the  \ka\ potential for the Heisenberg metric
(\ref{heismet}). Integrating equation 
\[
K'\,K'' = - 1
\]
we get
\[
K' = \sqrt{c - 2f}\ , \ \ \ K'' = -\frac{1}{\sqrt{c - 2f}}\ ,
\]
where $c$ is the integration constant which, without loss of
generality, we may set to zero.
Substituting these expressions into (\ref{metricII}) and ignoring an
overall constant factor we obtain the Heisenberg metric
(\ref{heist}). Note that the prefered complex structure with respect
to which the \ka\ potential is defined is the one whose associated
\ka\ form is $U(1)$-invariant. It is the two-form $\Omega_z$
(\ref{kahlerforms}) presented in section~\ref{heissect}.

\section{Exotic asymptotics: Bianchi types VII$_0$ and VI$_0$}
  \label{exoticsect} 

In this section we propose to investigate $p$-brane solutions whose
asymptotics are 
more unusual than the ones considered in section \ref{vacuumsect}. We
turn to  Bianchi types VII$_0$ and VI$_0$ whose groups of isometric
motions are $E(2)$ and $E(1,1)$ respectively (see
Table~\ref{bianchitable}).\footnote{We consider type VII$_0$ spaces
  before type VI$_0$ spaces because the type
  VII$_0$ metric is in some sense simpler since its isometry group is
  Euclidean.} 

We do not discuss the most general manifolds of the above types but
rather focus on self-dual metrics (which are \hk\ and hence BPS).
Vacuum four-metrics of this kind
are Ricci-flat \ka\ metrics. The \ein's equations
(\ref{r00})-(\ref{r33}) in the self-dual case reduce to a set of
first-order ODE's:
\begin{eqnarray}
\frac{2}{a}\, a' & = & -\, n_1\,a^2 + n_2\,b^2 + n_3\,c^2 \ , \label{sd1} \\
\frac{2}{b}\, b' & = & -\, n_2\,b^2 + n_3\,c^2 + n_1\,a^2 \ , \label{sd2} \\ 
\frac{2}{c}\, c' & = & -\, n_3\,c^2 + n_1\,a^2 + n_2\,b^2 \ . \label{sd3}
\end{eqnarray} 
For convenience we have introduced another radial coordinate
$\eta$ in place of $t$ such that $dt = abc\,
d\eta$ and $(\ )^{'}$ denotes differentiation with respect to $\eta$. 

In sections~\ref{sevenzero} and \ref{sixzero} we solve equations
(\ref{sd1})-(\ref{sd3}) to obtain self-dual vacuum Bianchi type
VII$_0$ and  VI$_0$
metrics and discuss their properties. In section~\ref{exoticadssect} we
discuss Bianchi  type VII$_0$ and VI$_0$ manifolds with negative
cosmological constant.

\subsection{Vacuum solutions of Bianchi type VII$_0$ and
  Solvmanifolds} \label{sevenzero} 

The group of isometries of a self-dual Bianchi type VII$_0$ metric is
$E(2)$, whose structure constants are $n_1 = n_2 = 1$ and $n_3 = 0$
and a set of left-invariant one-forms is:
\begin{eqnarray} \label{sevenzeroforms}
\sigma^1 & = & \cos\,\tau\, dx + \sin\,\tau\, dy\ , \\
\sigma^2 & = & -\, \sin\,\tau\,dx + \cos\,\tau\,dy\ , \nonumber\\
\sigma^3 & = & d\tau \ .\nonumber
\end{eqnarray} 
The self-duality equations (\ref{sd1})-(\ref{sd3}) become:
\begin{eqnarray}
\frac{2}{a}\, a' & = & -\,a^2 + b^2\ , \nonumber\\
\frac{2}{b}\, b' & = & -\,b^2 + a^2 \ , \nonumber \\ 
\frac{2}{c}\, c' & = & a^2 + b^2 \ . \nonumber
\end{eqnarray} 
These are easily solved to yield the metric:
\ben \label{sevenmet}
ds^2 = \frac{\lambda^2}{2}\, \sinh\,2\eta \left(d\,\eta^2 +
  (\sigma^3)^2\right) + \coth\,\eta\,(\sigma^1)^2+ \tanh\,
\eta\,(\sigma^2)^2\ , 
\een
where $\lambda$ is the integration constant.
Let us estimate the volume of a large metric ball as was done is
section~\ref{vacuumsect} for the Heisenberg manifold. Introducing as before
the effective radial coordinate $t$ to return to the ansatz
(\ref{bianchiansatz}), we find that the metric volume grows as $t^2$ for large
$t$. Interestingly, this is the predicted volume growth of another
type of a degeneration of the K3 surface
in the Kobayashi's review \cite{kobayashi}. In fact Kobayashi proved
the existence and completeness of a gravitational instanton whose
three-dimensional hypersurfaces $t = const$ represent a collapse of
{\em Solvmanifolds}.\footnote{Usually Bianchi type VI$_0$ group
  $E(1,1)$ is referred to as {\em Solv} or Solvable group. It is clear
  however that it is the type VII$_0$ manifold that has the volume
  growth predicted by Kobayashi. Its associated isometry group $E(2)$
  is also solvable.} The non-compact complete metric on a degeneration
of the K3 surface is expected to have quadratic volume growth of
large metric spheres and have as an asymptotic metric the standard
flat metric on $\co^* \times \co^*$. It is not known explicitly. The present
situation parallels the one we have already encountered with the
Heisenberg metric. The
Heisenberg metric (\ref{heismet}) is singular at the origin, but the
singularity is resolved by passing to another self-dual metric, the
BKTY gravitational instanton, whose asymptotic form is exponentially
close to  the
Heisenberg metric.  The singularity at the origin of the
Bianchi type VII$_0$ metric (\ref{sevenmet}) may be resolved by passing
to a non-singular manifold, whose existence and completeness is
guaranteed by the general theorem of Kobayashi \cite{kobayashi}. 

As we have pointed out, the large metric spheres have quadratic volume
growth for large $t$. According to the estimates in
section~\ref{actionsect}, this is the critical volume growth for which the
boundary term contribution to the gravitational action is constant and
finite.

Alternatively, metric (\ref{sevenmet}) may be obtained by solving the
\ma\  equation (\ref{maeqn}) for a Ricci-flat \ka\ metric
with appropriate
symmetries. If we assume that the \ka\ potential $K$ is independent of
the imaginary parts of $z^1 = u^1+ i\,v^1$ and $z^2 = u^2+ i\,v^2$ we
obtain a metric with two commuting holomorphic isometries. If one further
assumes that $K$ depends only on the combination $\sqrt{(u^1)^2 +(u^2)^2}$ one
gains an extra $SO(2)$ isometric action. From the first glance the
resulting metric appears to be invariant under the direct product
$SO(2)\times \re^2$ but in fact it turns out that the group of
isometries is the semi-direct product $SO(2)\ltimes \re^2\equiv E(2)$. Thus
one obtains a Bianchi type VII$_0$ metric. 

A short explicit calculation reveals that in polar coordinates $\{r,\tau\}$
\[
u^1 = r\, \cos\,\tau\ ,\ \ \ u^2 = r\, \sin\,\tau
\] 
the \ka\ potential depends only on $r$ and the metric becomes:
\ben \label{calabisevenmet}
ds^2 = K''\, dr^2 + r\,K' \, (\sigma^3)^2 +
\frac{K'}{r}\,\left((\sigma^1)^2 + (\sigma^2)^2\right) + \left(K''
- \frac{K'}{r}\right)\, (\sigma^1)^2\ ,
\een
where $K' = dK/dr$ and $\{\sigma^k\}$ are the left-invariant $E(2)$
one-forms given in (\ref{sevenzeroforms}). Then the \ma\ equation
reduces to an ODE first written down by Calabi \cite{calabi}:
\ben \label{calabieqn}
\frac{K''\, K'}{r} = e^{-\,\Lambda\, K}\ ,
\een
with $\Lambda$ the cosmological constant. When $\Lambda = 0$, the
vacuum case, Calabi found that 
\ben \label{kpotseven}
K'(r) = \sqrt{r^2 - a^2}\ ,
\een
where $a$ is an integration constant. The metric (\ref{calabisevenmet})
with (\ref{kpotseven}) is precisely the metric
(\ref{sevenmet}). Incidentally, this metric is the helicoid metric
found by Nutku \cite{nutku} who obtained it using the connection
between the real \ma\ equation and minimal surfaces. It is singular at
$\eta = 0$.

\subsection{Vacuum solutions of Bianchi type VI$_0$}
\label{sixzero} 

The group of motions preserving  Bianchi type VI$_0$ metrics is
$E(1,1)$. From Table~\ref{bianchitable} the structure constants are $n_1
= 1,\ n_2 = -1$ and $n_3 = 0$, and hence the 
left-invariant one-forms are:
\begin{eqnarray} \label{sixzeroforms}
\sigma^1 & = & \cosh\,\tau\, dx + \sinh\,\tau\, dy\ , \\
\sigma^2 & = & \sinh\,\tau\,dx + \cosh\,\tau\,dy\ , \nonumber\\
\sigma^3 & = & d\tau \ .\nonumber
\end{eqnarray} 
The self-duality equations (\ref{sd1})-(\ref{sd3}) become:
\begin{eqnarray}
\frac{2}{a}\, a' & = & -\,(a^2 + b^2)\ , \nonumber \\
\frac{2}{b}\, b' & = & a^2 + b^2 \ , \nonumber \\ 
\frac{2}{c}\, c' & = & a^2 - b^2 \ . \nonumber
\end{eqnarray}
These can be easily solved to give the following Ricci-flat \ka\
metric:
\ben \label{sixmet}
ds^2 = \frac{\lambda^2}{2}\, \sin\,2\eta \left(d\,\eta^2 +
  (\sigma^3)^2\right) + \cot\,\eta\,(\sigma^1)^2+ \tan\, \eta\,(\sigma^2)^2\ ,
\een
where $\lambda$ is the integration constant.
This metric is incomplete at the origin $\eta = 0$. 

Such self-dual metrics of Bianchi type VI$_0$ were also displayed by
Nutku \cite{nutku} and were referred to as catenoid metrics.

One can find a description of the metric (\ref{sixmet}) in terms
of a \ka\ potential as was done in the previous section for the type
VII$_0$ metric (\ref{sevenmet}).
Again assuming that the \ka\ potential $K$ is independent of
the imaginary parts of $z^1 = u^1+ i\,v^1$ and $z^2 = u^2+ i\,v^2$, we
obtain a metric with two commuting holomorphic isometries. Now 
assume that $K$ depends only on the combination $\sqrt{(u^1)^2
  -(u^2)^2}$ thus gaining an $SO(1,1)$ isometric action. The
resulting metric has as its group of
isometries the semi-direct product $SO(1,1)\ltimes \re^2\equiv E(1,1)$. 

Defining new coordinates $\{r,\tau\}$ as
\[
u^1 = r\, \cosh\,\tau\ ,\ \ \ u^2 = r\, \sinh\,\tau
\] 
we find that the \ka\ potential depends only on $r$ and the metric becomes:
\ben 
ds^2 = K''\, dr^2 - r\,K' \, (\sigma^3)^2 +
\frac{K'}{r}\,\left((\sigma^1)^2 + (\sigma^2)^2\right) + \left(K''
- \frac{K'}{r}\right)\, (\sigma^1)^2\ ,
\een
where $\{\sigma^k\}$ are the left-invariant $E(1,1)$
one-forms (\ref{sixzeroforms}). The \ma\ equation in this
case differs from Calabi's equation (\ref{calabieqn}) by a sign:
\ben \label{sixcalabieqn}
\frac{K''\, K'}{r} = - e^{-\,\Lambda\, K}\ .
\een
In the vacuum case $\Lambda = 0$ and the \ma\ equation
\[
\frac{K''\, K'}{r} = - 1
\] 
is solved by
\ben \label{kpotsix}
K'(r) = \sqrt{a^2 - r^2}\ ,
\een
where $a$ is the integration constant. This is the  metric
(\ref{sixmet}). 
 
\subsection{Bianchi type VII$_0$ and VI$_0$ with negative cosmological
  constant} \label{exoticadssect}
If the cosmological constant $\Lambda$ is negative, Calabi
\cite{calabi} proved that there exists a solution of (\ref{calabieqn}) 
giving a complete non-singular metric on $\re^4$. Unlike the analogous
metric of Bianchi type II (the Bergmann metric of
section~\ref{bergmannsect}), but like the Kasner--Anti-de Sitter metric
of section~\ref{kasnerads}, this metric is not homogeneous.  

Analogously, Calabi's argument concerning the
solution of equation (\ref{calabieqn}) with negative cosmological
constant is applicable to the Bianchi type VI$_0$ case. It may thus be
argued that 
solutions of (\ref{sixcalabieqn}) exist, although the completeness of
the metrics has to be demonstrated.

\section{Higher-dimensional examples of domain walls}  \label{highdimsect}

In this section we would like to give examples of domain walls
of the form $M \times \euc^{\, p-3,1}$ in eleven dimensions, where
manifold $M$ remains hypersurface homogeneous but 
has dimension higher than four. Firstly we describe Calabi--Yau
manifolds which are the higher-dimensional 
generalizations of the BKTY instanton of section~\ref{bktysect}. We
find their asymptotic metrics by solving the vacuum \ein's equations
in $2n$ dimensions. We then give particular examples of such
asymptotic metrics which arise as extensions of the vacuum Bianchi
type II or Heisenberg metric (\ref{heist}) to higher dimensions. In
addition we present three series of higher-dimensional
metrics originating from four-metrics of other Bianchi types: the
type I or Kasner (\ref{kasnermet}), the
type VII$_0$ (\ref{sevenmet}) and type VI$_0$ (\ref{sixmet})
metrics. We do so by generalising the relevant \ma\ equations and arguing the
existence of solutions which provide the \ka\ potentials for the
metrics in question.
By the extensions of Bianchi type metrics from four to
arbitrary number of dimensions we mean the following. The 
Bianchi type I isometry group $\re^3$ extended to $n+1$ dimensions is simply
$\re^n$. The Bianchi type II 
three-dimensional group of isometries given by the left-invariant
one-forms (\ref{heis1forms}) may be easily generalized to a
$(2n+1)$-dimensional group parametrized by $\{x_i,y_i,\tau\}$, where $i
= 1,\ldots,n$, with left-invariant one-forms:
\begin{eqnarray} \label{highheis1forms}
\sigma_i^1 & = & dx_i\ , \\
\sigma_i^2 & = & dy_i\ ,\nonumber \\
\sigma ^3 & = & d\tau-\sum_{i = 1}^n\,x_i\,dy_i\ ,\nonumber
\end{eqnarray}
satisfying
\[
d\sigma^3 = -\sum_i\, \sigma_i^1\wedge \sigma_i^2\ .
\]
The Bianchi type VI$_0$ and VII$_0$ groups are three-dimensional
groups $E(1,1)$ and $E(2)$ respectively. In higher dimensions these
become $E(n-1,1)$ and $E(n)$ respectively. 

\subsection{Higher-dimensional BKTY metrics and their asymptotic forms}

In this section we shall rely heavily on
the reference \cite{kobayashi} in which  Kobayashi proves the
existence  theorem for complete Ricci-flat \ka\
metrics on $X-D$ with $c_1(X) = [D]$, where $X$ is a Fano
manifold\footnote{$X$ is Fano if it has an ample canonical bundle, or,
in other words, if its first Chern class is positive, $c_1 >0$.} and
$D$ is a complex codimension one  hypersurface in $X$. Here $[D]$ is a
Poincar\'e dual of $D$. From Yau's solution to Calabi's conjecture one
may infer that $D$ carries a Ricci-flat \ka\ metric. Although the
gravitational instanton is not known explicitly, Kobayashi provides
some detailed information on the asymptotic form of the metric.  
It has the following properties.

Let $t(p)$ measure the distance from some fixed point in $X-D$ to
a point $p\in X-D$. Then far away from the chosen fixed point,
i.e. for large $t$, the metric spheres have a structure of an
$S^1$-bundle over $D$. The size of the fibre, with respect to the
induced metric on the metric spheres, decays as
$t^{-\frac{n-1}{n+1}}$, while the radius of the $(n-1)$
complex-dimensional base grows as
$t^\frac{1}{n+1}$. From this one estimates the volume
growth of a large metric ball to be:
\ben
Vol\sim \int t^{2(n-1)\cdot\frac{1}{n+1}}\cdot t^{-\frac{n-1}{n+1}} dt
\sim t^{\frac{2n}{n+1}}\ .
\een 

In this section we shall use the above information to make an ansatz
for the asymptotic form of the metric and to show that it is an exact
solution of the vacuum \ein's equations. We find that although the
solution is Ricci-flat and \ka\ it is singular. In fact, it bears the
same  relation
to the gravitational instantons of Kobayashi as does the Heisenberg
metric to the BKTY gravitational instanton. Outside a
compact set the complete metric differs from this asymptotic form by 
exponentially small terms.

It can be easily seen that the Heisenberg manifold (\ref{heist}) is
a special case of this setup for $n=2$. As we have already described
in section \ref{bktysect}, Kobayashi
points out that for $n=2$ these gravitational
instantons arise as certain degenerations of the $K3$ surface. The
metric spheres at large $t$ represent a collapse of a Nilmanifold to
a flat $T^2$, and the volume of a metric ball
grows as $t^{4/3}$. The Ricci-flat metric on $D$ is flat only in this
case. 

Consider the following ansatz compatible with the above remarks: 
\ben  \label{highdimansatz}
ds^2 = dt^2 +  a^2(t) g_{ab} dx^a dx^b\ + c^2(t)(d\tau - 2 A)^2\ .
\een 
Here $g_{ab}$ is the complete Ricci-flat \ka\ metric on $D$, $a,b = 1,
\cdots, 2(n-1)$; $\tau$ is the periodic coordinate on the canonical
bundle over $D$ and $A$ is a one-form that depends only on $x^a$ such
that its exterior
derivative is proportional to the \ka\ form on $D$, $dA = -\sigma J$,
$\sigma$ is constant.

The \ein's equations for (\ref{highdimansatz}) reduce to a system of
second order ODE's for the functions $a(t)$ and $c(t)$:
\begin{eqnarray}
0 & = & \frac{\ddot{a}}{a} + \frac{\dot{a}\dot{c}}{a c} + (2n-3)
\left(\frac{\dot{a}}{a} \right)^2 +  2 \sigma^2 \frac{c^2}{a^4}\ ,\label{highdim1} \\
0 & = & \frac{\ddot{c}}{c} + 2(n-1)\frac{\ddot{a}}{a}\ , \label{highdim2}\\
0 & = & \frac{\ddot{c}}{c} + 2(n-1) \frac{\dot{a}\dot{c}}{a c} -
2(n-1)\sigma^2 \frac{c^2}{a^4}\ . \label{highdim3}
\end{eqnarray}
 We therefore look for
solutions with polynomial dependence on the radial coordinate $t$ of the form:
\ben \label{powerlaw}
c(t) = \mu t^{\lambda_1}\ , a(t) = \nu t^{\lambda_2}\ ,
\een
where $\mu, \nu, \lambda_1, \lambda_2$ are constants. Substituting
(\ref{powerlaw}) into equations (\ref{highdim1})-(\ref{highdim3}) we find:
\begin{eqnarray}
0 & = & \lambda_2(\lambda_2 - 1) + \lambda_1 \lambda_2 + (2n-3)
\lambda_2^2 + 2\, \sigma^2\,  \frac{\mu^2}{\nu^4}\,
t^{2(1+\lambda_1-2\lambda_2)}\ ,\label{highdim4} \\
0 & = & \lambda_1(\lambda_1 - 1) + 2(n-1)\lambda_2(\lambda_2 - 1)\ ,
\label{highdim5} \\
0 & = & \lambda_1(\lambda_1 - 1) + 2(n-1)\lambda_1 \lambda_2
-2(n-1)\, \sigma^2\,  \frac{\mu^2}{\nu^4} t^{2(1+\lambda_1-2\lambda_2)}\ .
\label{highdim6}
\end{eqnarray}
In accordance with ansatz (\ref{powerlaw}) the equations
(\ref{highdim4})-(\ref{highdim6}) must reduce to algebraic equations
for the constants $\mu,\nu,\lambda_1,\lambda_2$. Hence we find that
$\lambda_1$ and $\lambda_2$ must satisfy:
\ben \label{rel1}
\lambda_1 = 2\lambda_2 - 1\ .
\een
Substituting (\ref{rel1}) into (\ref{highdim5}) we obtain a quadratic
equation for $\lambda_2$
\ben 
(n+1)\lambda_2^2 - (n+2)\lambda_2 +1 = 0\ ,
\een
which is solved by:
\ben
\lambda_2^{(1)} = \frac{1}{n+1}\ , \ \ \ \lambda_2^{(2)} = 1\ .
\een 
From (\ref{rel1}) we have:
\ben
\lambda_1^{(1)} = -\frac{n-1}{n+1}\ , \ \ \ \lambda_1^{(2)} = 1\ .
\een 
We discard the pair $(\lambda^{(2)}_1,\lambda^{(2)}_2) = (1,1)$ since
it satisfies both
(\ref{highdim4}) and (\ref{highdim6}) only for $n=0$. We are thus left with
the other pair of solutions $(\lambda_1,\lambda_2) \equiv
(\lambda^{(1)}_1,\lambda^{(1)}_2) =  (-\frac{n-1}{n+1},\frac{1}{n+1})$. 
Let us now find the constants $\mu, \nu$ and $\sigma$. From (\ref{highdim6}),
or equivalently from (\ref{highdim4}), we have
\ben \label{rel2}
\sigma^2\, \frac{\mu^2}{\nu^4} = \frac{1}{(n+1)^2}.
\een 
The values of $\mu$ and $\nu$ for $n=2$ may be read off from
the Heisenberg metric (\ref{heist}):  $\mu=(3/2)^{-1/3}$
and $\nu=(3/2)^{1/3}$. With these values (\ref{rel2}) gives 
\ben
\sigma = {1\over2}.
\een
Since parametrization of the one-form $A$ should not depend on the
dimension of $M$, we are compelled to choose constants $\mu$ and $\nu$
to  satisfy
(\ref{rel2})  with $\sigma = 1/2$ and consistent with their values for
$n=2$. An appropriate choice is:
\ben
\mu = \left(\frac{n+1}{2n}\right)^{-\frac{n-1}{n+1}}\ , \ \ \ \nu =
\left(\frac{n+1}{2n}\right)^{\frac{1}{n+1}}\ . 
\een 
Absorbing the constant $-\sigma = -1/2$ into the definition of the
one-form $A$ we may now write down the asymptotic metric for the
$2n$-dimensional  BKTY gravitational instanton:
\ben \label{highdimbkty}
ds^2 = dt^2 + \left(\frac{n+1}{2n} t \right)^{-2\frac{n-1}{n+1}} (d\tau +
A)^2 +  \left(\frac{n+1}{2n} t \right)^{\frac{2}{n+1}}g_{ab}dx^a dx^b\ ,
\een
where now $dA$ is precisely the \ka\ form on $D$.

Metric (\ref{highdimbkty}) indeed has the volume growth
predicted by Kobayashi:
\[
Vol\sim \int t^{-\frac{n-1}{n+1}} \cdot t^{2(n-1)\cdot \frac{1}{n+1}}
dt \sim t^{\frac{2n}{n+1}}\ .
\]  

\subsection{Bianchi type I}

Metrics of Kasner type (\ref{kasnermet}) exist in arbitrary number of
dimensions; the metric becomes:
\[
ds^2= dt^2 + t ^{2 \alpha_1} (dx^1)^2 + t ^{2 \alpha_2} (dx^2)^2+
\ldots + t^{2 \alpha_n} (dx^n)^2,
\]
where 
\ben \label{kasnerconstrhigh}
\sum_{k=1}^n \alpha_k= 1 = \sum_{k=1}^n \alpha_k^2\ .
\een
Just as the Kasner metric (\ref{kasnermet}), these metrics are not BPS.

\subsection{Bianchi type II}
A particular case of the asymptotic BKTY $2n$-dimensional metric
(\ref{highdimbkty}) is that whose isometry group is the
higher-dimensional Bianchi type II group. In this case the arbitrary
Calabi--Yau $(2n-2)$-dimensional metric $g_{ab}dx^a\,dx^b$ is the flat
metric 
\[
\sum_{i=1}^{n-1}\, (\sigma_i^1)^2 + (\sigma_i^2)^2 \ .
\]
and the term involving the connection on the canonical bundle over
$g_{ab}dx^a\,dx^b$ is simply $(\sigma^3)^2$; where $\{\sigma_i^1,
\sigma_i^2, \sigma^3\}$ are the one-forms defined in
(\ref{highheis1forms}).  

To find the \ka\ potential generating the metric (\ref{highdimbkty})
for the special case of Bianchi type II isometry group we solve
the \ma\ equation (\ref{maeqnIIhigh}) with $\Lambda = 0$:
\ben \label{maeqnhighbkty}
(K')^{n-1}\,K'' = (-1)^{n-1}\ .
\een
 It is
sufficient to know $K'(f)$ where $f = z+\bar{z} - \sum_i |w^i|^2$
since the higher-dimensional Bianchi type II metric is expressed in
terms of $K'$ and $K''$ as follows:
\[
ds^2 = -K'\,\sum_i\,dw^i\wedge d\bar{w}^i + K''\,(dz -
  \sum_i\,\bar{w}^i\,dw^i)(d\bar{z} -  w^i\,d\bar{w}^i)\ ,
\]
which after coordinate redefinitions (\ref{bergmap}) with $\{w,x,y\}$
replaced by $\{w^i, x^i,y^i\}$ becomes
\ben \label{potIImet}
ds^2 = -K'\,\sum_i\left((dx^i)^2 +(dy^i)^2\right) +  K''\,df^2 +
K''\,(d\tau - \sum_i\,x^i\,dy^i)^2\ .
\een
Integrating equation (\ref{maeqnhighbkty}) we find:
\[
K' = (-1)^{\frac{n-1}{n}}\,(n\,f)^{1\over n}
\]
and hence
\[
K'' = (-1)^{\frac{n-1}{n}}\,(n\,f)^{-\frac{n-1}{n}}\ .
\]
To compare with the metric we have obtained by solving the
self-duality equations let us define a new radial coordinate $t$ such that
\[
f = \left(\frac{n+1}{2n}\,t\right)^{\frac{2n}{n+1}}\ .
\]
Written in terms of $t$ the metric (\ref{potIImet}) is the same as
(\ref{highdimbkty}) up to an overall constant factor.

\subsection{Bianchi type VII$_0$ and VI$_0$}

The analysis based on solving the \ma\ equations (\ref{calabieqn}) and
(\ref{sixcalabieqn}) may be extended to
arbitrary number of dimensions. In fact in the case of Bianchi type
VII$_0$ it was done by Calabi in
\cite{calabi}. If $M$ has real dimension $2n$ with complex
coordinates $z^a,\ a = 1,\ldots, n$ and one assumes, as was done
in section~\ref{sevenzero}, that the \ka\ potential $K$ is independent of
the imaginary parts of $z^a = u^a + i\,v^a$ and is solely a function
of $r \equiv \sqrt{(u^1)^2+\ldots + (u^n)^2}$, the resulting metric will have
the isometry group $E(n)$. The \ma\ equation reduces to 
\[
\left(\frac{K'}{r}\right)^{n-1}\, K'' = 1\ .
\] 
It is solved by
\[
K(r) = \int_0^r (c + r^{1\over n})\, dr\ ,\ \ \ \ c = const\ .
\]
The manifold is a higher-dimensional vacuum Bianchi type VII$_0$ metric
whose isometry group is $E(n)$. It is incomplete.
Arguments analogous to the ones just given extend to the Bianchi type
VI$_0$ metrics. 

\section{Other Bianchi types: Bianchi type III} \label{bianchiIIIsect}

We have not attempted here to survey all known cohomogeneity one \ein\
metrics. Even in four dimensions this would be a formidable task. Some
pertinent references in that case are \cite{siklos,lorentz}. However
we would like to comment on the Bianchi type III situation since it may well
prove relevant for various applications of Anti--de Sitter
space-time. 

The most general diagonal Lorentzian Bianchi type III local solution is
given in \cite{maccullum}. A simple analytic continuation of the
metric in \cite{maccullum} gives a Riemannian metric with negative
scalar curvature $\Lambda<0$
\ben \label{bianchiIII}
ds^2 = \frac{3}{\Lambda}\, \left(\frac{d\tau^2}{\sinh^2\tau} +
  \frac{d\Omega_{-1}^2}{\sinh^2\tau} +
  \frac{d\alpha^2}{\tanh^2\tau}\right)\ ,
\een
 which is presumably the most general local solution
with this signature. 
Setting 
\[
t = \log\tanh \frac{\tau}{2}
\]
gives
\[
ds^2 = \frac{3}{\Lambda}\, \left(dt^2 + \sinh^2\,t\,d\Omega_{-1}^2 +
  \cosh^2\,t\, d\alpha^2\right)\ . 
\]
In (\ref{bianchiIII}) $d\Omega_{-1}^2$ is the standard metric on
${\cal H}^2$. The isometry group of the manifold is therefore
$SO(2,1)\times SO(2)$. The group $SO(2,1)$ has a two-dimensional
subgroup $\tilde{G}_2$ which acts transitively on ${\cal H}^2$ and
combined with $SO(2)$ we get a three-dimensional Lie group with
three-dimensional orbits whose Lie algebra corresponds to Bianchi type
III.
Explicitly we consider ${\cal H}^2$ in horospheric coordinates $(x,y)$
\[
\frac{dx^2 + dy^2}{y^2}
\]
and $\tilde{G}_2 = \re \ltimes \re$ is  generated by $\partial/\partial\,x$ and
$x\,\partial/\partial\,x + y\,\partial/\partial\,y$.

In fact the metric (\ref{bianchiIII}) is that of hyperbolic four-space
${\cal H}^4$ (cf section~\ref{horospheresect}). This may be seen by
isometrically embedding (\ref{bianchiIII}) into $\euc^{4,1}$ as:
\[
(X^0)^2 - (X^1)^2- (X^2)^2- (X^3)^2- (X^4)^2 = 1\ ,
\]
where 
\begin{eqnarray}
X^3 & = & \cosh\,t\, \cos\alpha\ , \nonumber \\
X^4 & = & \cosh\,t\, \sin\alpha\ , \nonumber \\
X^0 + X^1 &=& \frac{1}{y}\,\sinh\,t\ ,\nonumber \\
X^0 - X^1 &=& \left(y + \frac{x^2}{y} \right)\,\sinh\,t\ ,\nonumber \\
X^2 &=& \frac{x}{y}\,\sinh\,t\ .\nonumber
\end{eqnarray}
It is now obvious that the Bianchi type III solution
(\ref{bianchiIII}) may be extended to $(n+2)$ dimensions by replacing
the metric on ${\cal H}^2$ by that on ${\cal H}^n$. The group
$\tilde{G}_2$ is replaced by the group $\tilde{G}_n =
\re\ltimes\re^{n-1}$ generated by $\partial/\partial\,x^i$ and
$x^i\,\partial/\partial\,x^i + y\,\partial/\partial\,y$, $i =
1,\ldots, n-1$. then the generalization of the Bianchi type III group
is $(\re\ltimes\re^{n-1})\times SO(2)$.

\section{Conclusions} \label{conc}

In this paper we have studied various solutions of M-theory having the
symmetries of a domain wall. Our most important example
(\ref{heismet}) is  BPS and is
based on the Bianchi type II group, otherwise known as $Nil$ or
Heisenberg. Usually this is regarded as an orbifold solution with
singularity. We have shown how the singularity may be resolved to give
a complete non-singular solution representing a single-sided domain
wall. We have also shown how this example may be generalized to higher
dimensions.

Finally, we have considered a number of related solutions, some BPS,
both in four and higher dimensions which we believe may be relevant
to, for example, the AdS/CFT correspondence and other future
applications of eleven-dimensional supergravity.

\end{document}